\documentclass[runningheads]{llncs}

\usepackage[utf8]{inputenc}
\usepackage[T1]{fontenc}
\usepackage{graphicx}
\usepackage{dcolumn}
\usepackage{bm}
\usepackage{hyperref}
\usepackage{braket}
\usepackage{amsmath}
\hypersetup{
    colorlinks=true,
    linkcolor=blue,
    filecolor=magenta,
    urlcolor=cyan,
    citecolor=blue,
}
\usepackage{cite}
\usepackage{amssymb}
\usepackage{booktabs}
\usepackage{xcolor}
\usepackage{url}
\usepackage{float}
\usepackage{caption}
\title{SPSA Hyperparameter Tuning for Variational Quantum Natural Language Inference}
\titlerunning{SPSA Tuning for Variational QNLI}

\author{
Nayan D'Souza\inst{1,2} \and
Christopher J. Agostino\inst{1}
}

\authorrunning{N. D'Souza and C. Agostino}

\institute{
NPC Worldwide, Bloomington, IN, USA \\
\email{cjp.agostino@gmail.com} \and
Department of Linguistics, Indiana University, Bloomington, IN, USA
}

\begin{document}
\maketitle

\begin{abstract}
Training variational quantum models requires choosing between parameter-shift gradients, which are exact but cost $\mathcal{O}(P)$ forward evaluations, and simultaneous perturbation stochastic approximation (SPSA), which uses only two samples but produces high-variance estimates that can degrade optimisation on small supervised tasks. Whether the cheap gradient is usable depends on the variance that results from different choices of the SPSA perturbation scale, learning rate, and gain-decay schedule. We varied those quantities across a broad grid on a 6-qubit, 60-parameter QNLI classifier and compared the best configurations to parameter-shift AdamW and BuresQNG. AdamW-style SPSA with $c_0=0.01$, $\eta=0.10$, $\gamma=0.10$ reached $55\% \pm 11\%$ test accuracy, improving over the default configuration ($49\% \pm 6\%$) but remaining 16--19 percentage points below the parameter-shift baselines because the two-sample SPSA gradient estimate has too much variance for reliable optimisation of 60 parameters in 40 epochs. Classical-gain SPSA and Bures-preconditioned SPSA performed worse, at $51\%$ and $46\%$ respectively. Bures-preconditioning a noisy two-sample SPSA gradient amplifies perturbation noise.

\keywords{SPSA, parameter-shift rule, quantum natural gradient, variational quantum algorithms, QNLP}
\end{abstract}

\section{Introduction}

In 1951 Herbert Robbins and Sutton Monro showed that a stochastic process could find the root of a regression function without ever evaluating that function exactly, launching the field of stochastic approximation \cite{robbins1951}, and the following year Jack Kiefer and Jacob Wolfowitz transferred the stochastic approximation principle to gradient estimation, taking finite-difference measurements along randomly chosen coordinates from noisy observations \cite{kiefer1952} so the gradient estimate itself contained error and partial information. Stochastic gradient descent and related optimizers are the standard toolkit in modern machine learning, and quantum machine learning has imported these routines to tune parameterized quantum circuits. Whether real-valued optimisation methods are adequate for parameters whose effect is to modify complex Hilbert-space amplitudes is an open empirical question.

In the NISQ era, variational quantum algorithms are trained by querying noisy intermediate-scale devices repeatedly \cite{preskill2018}. They train parameterized quantum circuits by gradient descent, and the dominant expense is often the gradient estimate itself \cite{mari2020,banchi2020,schuld2021book}. Parameter-shift methods give exact gradients for many parameterized gates but require $\mathcal{O}(P)$ forward evaluations per parameter vector, where $P$ is the number of parameters \cite{mitarai2018,schuld2018evaluating,wierichs2020}. This cost is modest for small circuits but dominates wall-clock training time for larger encoders. Simultaneous perturbation stochastic approximation (SPSA), introduced by Spall four decades after Robbins-Monro, estimates the gradient with only two function evaluations regardless of $P$ \cite{spall1992,spall1998,spall2000}, though its two-sample estimate is high variance. The cost and variance of these estimators determine how much wall-clock time and gradient quality each optimization step provides, at a scale where trainability is already constrained by barren plateaus and exponentially flat landscapes \cite{mcclean2018,arrasmith2021,cerezo2021}.

Few studies compare exact-gradient optimisers and SPSA on the same supervised task, so the crossover point where SPSA becomes worthwhile is unclear. This gap is pronounced for supervised natural-language tasks, where the objective is non-convex and the readout layer couples circuit parameters to a classical classifier. Quantum natural language processing has so far been explored mainly through compositional categorical models and small-scale proof-of-concept classifiers \cite{coecke2010,coecke2020,meichanetzidis2020,lorenz2023}, and the optimization behaviour of these models under approximate gradients has not been studied as closely. In recent work on a small QNLI model, parameter-shift AdamW and diagonal Bures-metric quantum natural gradient (BuresQNG) reached roughly 72--74\% test accuracy while SPSA and Bures-preconditioned SPSA gave 49\% and 37\%, respectively. Two questions follow. First, was SPSA merely mis-tuned, or is its two-sample gradient estimate too noisy for reliable optimisation at this scale? Second, does the Bures metric, which helps when paired with exact parameter-shift gradients \cite{gomez2021,bowles2024}, help when applied to a noisy SPSA gradient, or does it instead magnify noise along sensitive circuit directions?

We grid-search SPSA hyperparameters on the same small QNLI task and compare classical-gain SPSA, AdamW-style SPSA, and Bures-preconditioned SPSA against parameter-shift AdamW and BuresQNG.

\section{Methods}

All experiments use the shared \texttt{QNLPModel} implemented in Qiskit with NumPy statevector simulation.
The circuit has $n_q=6$ qubits and $n_l=3$ layers.
Each layer applies an $R_x$ and an $R_z$ rotation to every qubit, followed by nearest-neighbor CNOTs, giving $n_\text{circuit}=2 n_q n_l = 36$ circuit parameters.
The readout head has $n_r=3$ bins and contributes 24 additional trainable weights and biases, for a total of 60 trainable parameters.

Sentence encoding is word-level.
The first six words of a sentence each initialise one qubit via fixed $R_y$ and $R_z$ rotations derived from a hash of the word.
Trainable layers then transform the state, and the readout computes a 3-bin probability distribution for each sentence.
For a premise--hypothesis pair, the model forms the concatenated feature vector $[\mathbf{p}, \mathbf{h}, \mathrm{sim}(\mathbf{p},\mathbf{h})]$, where $\mathbf{p}$ and $\mathbf{h}$ are the readout distributions and $\mathrm{sim}$ is their cosine similarity, and passes it through a linear classifier.

The dataset consists of 300 QNLI-style examples generated from 15 sentence templates with synonym substitution and three labels: entailment (0), contradiction (1), and neutral (2).
Examples are split 70/30 (210 train / 90 test), and all optimizers run for 40 epochs with cross-entropy loss.

We compare five optimizers:
\begin{enumerate}
\item AdamW with parameter-shift gradients, learning rate $0.2$, $\beta_1=0.9$, $\beta_2=0.999$, weight decay $0.01$ \cite{kingma2014,loshchilov2017}.
\item BuresQNG with parameter-shift gradients, learning rate $0.2$, diagonal Bures metric recomputed every 5 epochs, AdamW-style moment update.
\item SPSA-classical: raw SPSA update $\boldsymbol{\theta}_{k+1} = \boldsymbol{\theta}_k - a_k \hat{g}_k$ with $a_k = \eta / k^\gamma$.
\item SPSA-AdamW: the SPSA gradient estimate $\hat{g}_k$ is fed into AdamW-style first- and second-moment accumulation.
\item BuresSPSA: classical-gain SPSA preconditioned by the diagonal Bures metric on the 36 circuit parameters.
\end{enumerate}

The SPSA grid covers
\begin{equation}
c_0 \in \{0.01, 0.05, 0.10, 0.20\},\quad
\eta \in \{0.05, 0.10, 0.20, 0.50\},\quad
\gamma \in \{0.10, 0.30, 0.50\},
\end{equation}
giving 48 configurations per variant and 144 SPSA runs in total.
Each configuration is evaluated across random seeds 42, 2024, and 7, with perturbation scale $c_k = c_0 / k^\gamma$.
For each run we record training loss per epoch, final test loss, and test accuracy; aggregate results report mean $\pm$ standard deviation across seeds.

Wall-clock encoder times are approximately $4$~s per epoch for SPSA variants versus $2.5$~s for parameter-shift AdamW/BuresQNG.
Because the full forward pass dominates, the total training time of an SPSA epoch is comparable to or slightly longer than a parameter-shift epoch on this 60-parameter model; the expected $\mathcal{O}(P)$ savings do not materialise here.

\section{Results}

The best SPSA configurations and parameter-shift baselines are listed in Table~\ref{tab:summary}.
AdamW-style SPSA accounts for the first four ranks and nine of the top ten configurations.
Its best setting ($c_0=0.01$, $\eta=0.10$, $\gamma=0.10$) gives $55\% \pm 11\%$ test accuracy and $1.0 \pm 0.02$ test loss, compared with $49\% \pm 6\%$ for the default SPSA configuration and roughly 72--74\% for AdamW and BuresQNG.

Classical-gain SPSA gives $51\% \pm 8\%$ at best, with $\eta=0.50$ and $\gamma=0.50$.
The best BuresSPSA configuration reaches $46\%$, and most BuresSPSA settings fall between $30\%$ and $40\%$.

\begin{table}[ht]
\centering
\caption{Top grid configurations and parameter-shift baselines. Test accuracy and loss are mean $\pm$ std across three seeds.}
\label{tab:summary}
\small
\begin{tabular}{@{}lcccccc@{}}
\toprule
Optimizer / Variant & $c_0$ & $\eta$ & $\gamma$ & Test Acc & Test Loss \\
\midrule
AdamW baseline & --- & 0.2 & --- & 72\% $\pm$ 13\% & 0.60 $\pm$ 0.05 \\
BuresQNG baseline & --- & 0.2 & --- & 74\% $\pm$ 9\% & 0.60 $\pm$ 0.06 \\[2pt]
spsa\_adamw & 0.01 & 0.10 & 0.10 & 55\% $\pm$ 11\% & 1.0 $\pm$ 0.02 \\
spsa\_adamw & 0.05 & 0.10 & 0.10 & 54\% $\pm$ 12\% & 1.0 $\pm$ 0.03 \\
spsa\_adamw & 0.20 & 0.05 & 0.30 & 53\% $\pm$ 16\% & 1.0 $\pm$ 0.05 \\
spsa\_adamw & 0.10 & 0.10 & 0.10 & 52\% $\pm$ 7\% & 1.1 $\pm$ 0.01 \\
spsa\_classical & 0.10 & 0.50 & 0.50 & 51\% $\pm$ 8\% & 1.1 $\pm$ 0.02 \\[2pt]
SPSA default (Exp.~12) & 0.05 & 0.2 & 0.10 & 49\% $\pm$ 6\% & 1.1 $\pm$ 0.05 \\
BuresSPSA default (Exp.~12) & 0.05 & 0.2 & 0.10 & 37\% $\pm$ 11\% & 1.1 $\pm$ 0.03 \\
Best BuresSPSA & 0.05 & 0.50 & 0.50 & 46\% & --- \\
\bottomrule
\end{tabular}
\end{table}

\begin{figure}[ht]
\centering
\includegraphics[width=0.95\columnwidth]{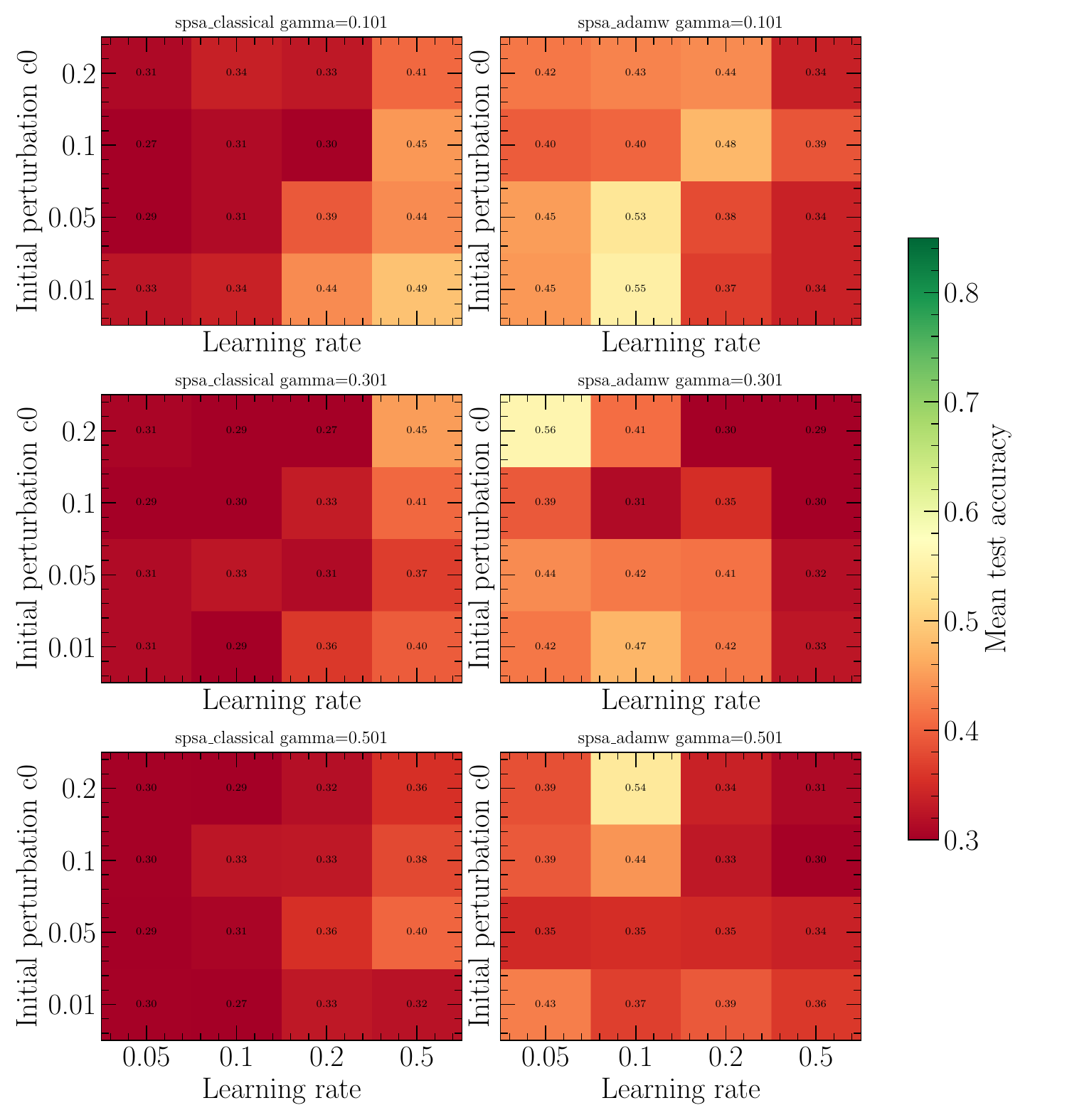}
\caption{Test accuracy heatmap over the SPSA grid. AdamW-style SPSA (top row) ranks above the other variants, but no configuration approaches the parameter-shift baselines.
\label{fig:heatmap}}
\end{figure}

In the full grid (Figure~\ref{fig:heatmap}), AdamW-style SPSA accounts for nine of the top ten configurations.
The best SPSA settings cluster at small perturbations ($c_0 \le 0.05$) and slow decay ($\gamma=0.10$), with moderate learning rates around $0.10$.
Large learning rates ($0.50$) frequently cause divergence or high variance; SPSA-AdamW with $\eta=0.50$ and $\gamma=0.10$ gives $31\%$ test accuracy.
BuresSPSA is both low-performing and unstable, with standard deviations across seeds often 10--18 percentage points.
Test loss stays near or above $1.0$ for all SPSA variants, and AdamW and BuresQNG both settle near $0.60$.

\begin{figure}[ht]
\centering
\includegraphics[width=0.95\columnwidth]{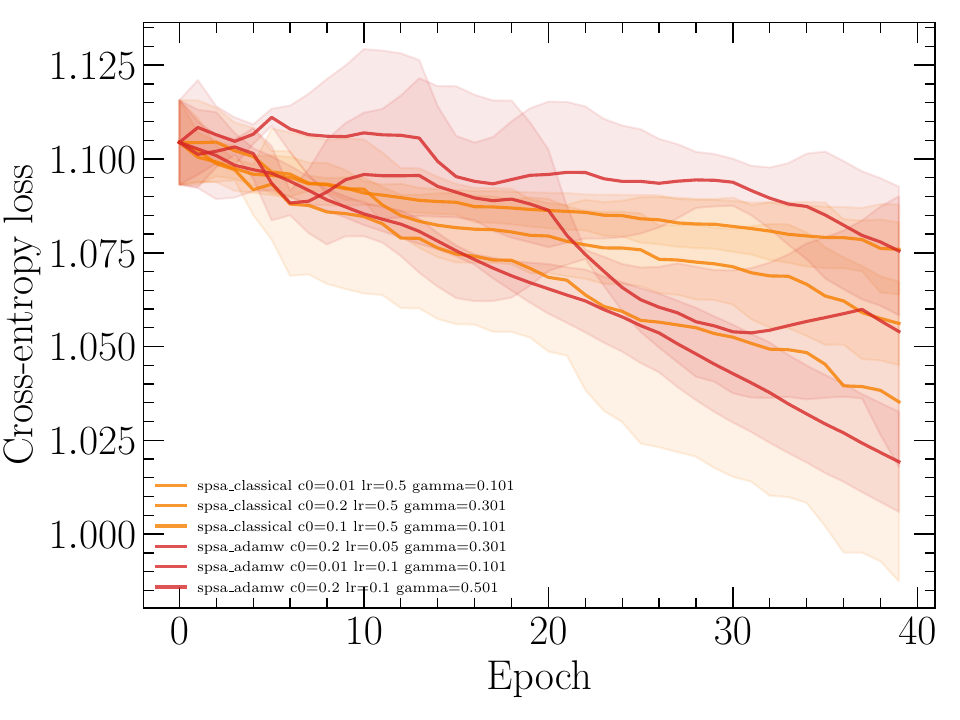}
\caption{Training-loss convergence for top configurations and parameter-shift baselines. AdamW-style SPSA configs converge to higher loss than parameter-shift optimizers, and BuresSPSA remains unstable.
\label{fig:convergence}}
\end{figure}

Training-loss curves for selected top configurations are shown in Figure~\ref{fig:convergence}.
AdamW and BuresQNG descend rapidly to a stable cross-entropy near $0.60$.
AdamW-style SPSA configurations stop descending near $1.0$, while BuresSPSA remains unstable and often increases loss after an initial drop.

Even the best SPSA-AdamW configuration has a standard deviation of $11$ percentage points across three seeds.
Several higher-accuracy configurations have much higher variance; $c_0=0.05$, $\eta=0.10$, $\gamma=0.50$ gives $50\% \pm 20\%$.
AdamW and BuresQNG have standard deviations of 13 and 9.5 percentage points at much higher mean accuracy.

\section{Discussion}

A broad grid search improves the best SPSA accuracy from $49\%$ to $55\%$, but the gap to parameter-shift AdamW/BuresQNG ($72$--$74\%$) stays at 16--19 percentage points.
The best settings cluster at small perturbations, moderate learning rate, and slow decay, without closing the gap.
The two-sample SPSA gradient estimate has enough variance to explain the gap.
With only 60 parameters, parameter-shift obtains a much cleaner gradient at acceptable cost, while SPSA's variance cannot be fully tamed by AdamW moments or by the Bures metric.

The diagonal Bures metric is computed from exact statevectors and reflects the local geometry of the quantum state space.
With parameter-shift gradients it rescales updates along that geometry; with SPSA gradients it rescales noisy estimates along the same sensitive directions.
Because the metric rescales each parameter by the inverse of its quantum sensitivity, directions that are geometrically steep receive larger updates; if the SPSA estimate has high variance along those directions, the preconditioner amplifies the error.
For SPSA to benefit from geometric preconditioning, the gradient variance would need to be reduced first, for example by averaging multiple simultaneous perturbations.

AdamW-style moment accumulation consistently outperforms the classical decaying-gain schedule.
The first moment smooths the noisy gradient direction and the second moment adapts step sizes per parameter.
That smoothing cannot remove the variance of the underlying estimate; the best SPSA-AdamW accuracy is still far below parameter-shift AdamW, and its standard deviation across seeds remains high.

On much larger encoders the wall-clock cost of parameter-shift could make SPSA attractive despite its noisier gradient, provided the variance is reduced through averaged perturbations or a warm-start to parameter-shift fine-tuning \cite{kubler2020,harrow2021}. That regime is not the one studied here.

\section{Conclusions}

We investigated whether the poor performance of SPSA on a small variational QNLI classifier could be remedied by hyperparameter tuning, and how the best SPSA configurations compare to parameter-shift AdamW and diagonal Bures-metric quantum natural gradient (BuresQNG). We grid-searched the initial perturbation scale $c_0$, the learning rate, and the perturbation decay across classical-gain, AdamW-style, and Bures-preconditioned SPSA on a 6-qubit, 60-parameter circuit trained on 300 QNLI examples. The results can be summarized as follows:

\begin{enumerate}
\item Hyperparameter tuning improves SPSA but does not close the gap. AdamW-style SPSA with $c_0=0.01$, $\eta=0.10$, $\gamma=0.10$ reaches $55\% \pm 11\%$ test accuracy, up from $49\% \pm 6\%$ for the default configuration, but still 16--19 percentage points below parameter-shift AdamW/BuresQNG.

\item AdamW-style momentum stabilises SPSA better than a classical decaying gain. AdamW-style SPSA accounts for nine of the top ten grid configurations, while classical-gain SPSA peaks at $51\%$ and requires a large learning rate ($\eta=0.50$) with slow decay ($\gamma=0.50$).

\item Bures preconditioning of a noisy SPSA gradient is counterproductive. BuresSPSA reaches at best $46\%$ and is unstable, with standard deviations across seeds often 10--18 percentage points, because the metric rescales high-variance estimates along sensitive circuit directions.

\item The variance of the two-sample SPSA estimate is the limiting factor. With only 60 parameters, parameter-shift provides a much cleaner gradient at acceptable cost, and neither AdamW moments nor the Bures metric can remove enough variance for reliable supervised learning in 40 epochs.
\end{enumerate}

SPSA could still become competitive on much larger encoders, where the wall-clock savings of $\mathcal{O}(1)$ gradient estimates matter, or when combined with variance reduction or warm-start strategies. That regime is not covered by this experiment.

\bibliographystyle{splncs04}
\bibliography{exp12b_refs}

\end{document}